\documentclass[12pt]{article}
\usepackage{graphicx}
\usepackage{hyperref}
\usepackage{amssymb}

\makeatletter
\renewcommand\section{\@startsection {section}{1}{\z@}%
                                   {-3.5ex \@plus -1ex \@minus -.2ex}
                                   {2.3ex \@plus.2ex}%
                                   {\normalfont\large\bfseries}}
\renewcommand\subsection{\@startsection{subsection}{2}{\z@}%
                                     {-3.25ex\@plus -1ex \@minus -.2ex}%
                                     {1.5ex \@plus .2ex}%
                                     {\normalfont\bfseries}}

\makeatother

\newcommand{\bea}{\begin{eqnarray}}
\newcommand{\eea}{\end{eqnarray}}
\newcommand{\be}{\begin{equation}}
\newcommand{\ee}{\end{equation}}

\newcommand{\del}{\partial}
\newcommand{\delbar}{\overline{\partial}}
\newcommand{\zbar}{\overline{z}}
\newcommand{\wbar}{\overline{w}}

\newcommand{\Z}{{\mathbb Z}}
\newcommand{\R}{{\mathbb R}}

\def\taubar{\overline{\tau}}
\def\Tr{{\rm Tr}}
\def\qhat{\hat{q}_0}

\begin{document}
\begin{titlepage}
\begin{center}

\today
\hfill                  hep-th/0608059

\hfill ITFA-06-29

\vskip 2 cm {\Large \bf A Farey Tail for Attractor Black Holes}
\vskip 1.25 cm { Jan de Boer\footnote{
jdeboer@science.uva.nl},
Miranda C.N. Cheng\footnote{
mcheng@science.uva.nl}, Robbert
Dijkgraaf\footnote{
rhd@science.uva.nl},\\[1mm] Jan Manschot\footnote{
manschot@science.uva.nl} and Erik Verlinde\footnote{
erikv@science.uva.nl} }\\
{\vskip 0.5cm  Institute for Theoretical Physics\\ University of Amsterdam\\
Valckenierstraat 65\\
1018 XE, Amsterdam\\
The Netherlands\\}

\end{center}

\vskip 2 cm

\begin{abstract}
\baselineskip=18pt

The microstates of 4d BPS black holes in IIA string theory
compactified on a Calabi-Yau manifold are counted by a
(generalized) elliptic genus of a (0,4) conformal field theory. By
exploiting a spectral flow that relates states with different
charges, and using the Rademacher formula, we find that the
elliptic genus has an exact asymptotic expansion in terms of
semi-classical saddle-points of the dual supergravity theory. This
generalizes the known "Black Hole Farey Tail" of
\cite{Dijkgraaf:2000fq} to the case of attractor black holes.

\end{abstract}

\end{titlepage}

\tableofcontents
\addtocontents{toc}{\protect\setcounter{tocdepth}{2}}

\pagestyle{plain}
\baselineskip=19pt

\section{Introduction}

One of the main successes of string theory has been the
microscopic explanation of black hole entropy. The microstates
for extremal BPS black holes are well understood in
theories with 16 or more supercharges. This includes the
original D1-D5-P system in type IIB theory on \(K3\times S^1\) for
which the microstates are represented by the elliptic genus of a
(4,4) CFT with target space given by a symmetric product of $K3$ \cite{Strominger:1996sh, Dijkgraaf:1996xw}.
The elliptic genus for this target space can be explicitly
computed, leading to a concrete and exact expression for the
number of BPS-states.

The D1-D5-P system has a well understood dual
description in terms of type IIB theory on $K3\times AdS_3\times
S^3$. A rather remarkable result, known as the Rademacher
series, is that the elliptic genus has an exact asymptotic
expansion, which has a natural interpretation as a sum over
semi-classical contributions of saddle-point configurations of the
dual supergravity theory.  This exact asymptotic expansion, together with its
semi-classical interpretation, has been coined
 the Black Hole Farey Tail \cite{Dijkgraaf:2000fq}.
Although the Farey tail was first introduced in the context of the D1-D5 system,
it applies to any system that has a microscopic
description in terms of a (decoupled) 2d conformal field theory
and has a dual description as a string/supergravity theory on a
spacetime that contains an asymptotically $AdS_3$.

The aim of this paper is to apply the Rademacher formula to black
holes in theories with eight supercharges and in this way extend
the Farey Tail to N=2 (or attractor) black holes.  Specifically, we
consider M-theory compactified on a Calabi-Yau three-fold $X$ and
study the supersymmetric bound states of wrapped M5-branes with
M2-branes. These states correspond to extremal four dimensional black holes
after further reduction on a circle. For this situation a
microscopic description was proposed quite a
while ago by Maldacena, Strominger and Witten (MSW)
\cite{Maldacena:1997de}, who showed that the black hole microstates are
represented by the supersymmetric ground states of a (0,4) conformal field theory.
These states are counted by an appropriately defined elliptic genus of the (0,4) CFT.

The interest in attractor black holes has been revived in recent
years due to the connection with topological string theory
discovered in \cite{LopesCardoso:1998wt, Ooguri:2004zv} and
subsequently studied by many different authors.
In particular, it was conjectured by Ooguri, Strominger and Vafa
(OSV) in \cite{Ooguri:2004zv} that the mixed partition function of
4d BPS black holes is equal to the absolute value squared of the
topological string partition function. Earlier, in a separate
development, a different connection between BPS states and
topological strings was discovered by Gopakumar and Vafa (GV)
\cite{Gopakumar:1998ii}, who showed that topological string theory
computes the number of five-dimensional BPS-invariants of wrapped
M2 branes in M-theory on a Calabi-Yau. The GV-result differs from
the OSV-conjecture in the sense that the topological string
coupling constant appears in an S-dual way. Recently, this fact
and the OSV conjecture have been considerably clarified in the
work of Gaiotto, Strominger, and Yin \cite{Gaiotto:2006ns}. These
authors used the CFT approach of MSW to show that the elliptic
genus of the (0,4) CFT has a low temperature expansion which
(approximately) looks like the square of the GV-partition
function. The OSV conjecture then follows from the modular
invariance of the elliptic genus, which at the same time naturally
explains the different appearances of the coupling constant.

In this paper we will show that elliptic genus of the (0,4) SCFT
can be written as a Rademacher series (or Farey tail expansion)
similar to that of the previously studied (4,4)
case.\footnote{Some related results were obtained independently in
\cite{Denef-Moore,Gaiotto:2006,Kraus:2006nb}.}  An important
property of the SCFT is the presence of a spectral flow that
relates states with different charges, and implies that the
elliptic genus can be expanded in terms of  theta functions. These
theta functions signal the presence of a set of chiral scalars in
the SCFT, while from a spacetime point of view their appearance
naturally follows from the Chern-Simons term in the effective
action.  We find that the Farey tail expansion contains subleading
contributions to each saddle point that can be interpreted as
being due to a virtual cloud of BPS-particles (actually, wrapped
M2-branes) that are "light" enough so that by themselves they do
not form a black hole. The degeneracies of these particles are, in
the large central charge limit, given in terms of the
Gopakumar-Vafa invariants. In this way we see that the results of
\cite{Gaiotto:2006ns} naturally fit in and to a certain extent
follow from our Attractor Farey Tail.

The outline of the paper is as follows: in section 2 we review the
bound states of wrapped M5 and M2 branes in M-theory on a
Calabi-Yau three-fold and explain the emergence of the spectral
flow. We then discuss the decoupling limit and the near horizon
geometry and describe the dimensionally reduced effective action
on $AdS_3$. In section 3 we turn to the M5 brane world volume
theory and its reduction to the (0,4) SCFT. Here we also define
the generalized elliptic genus which will give us the graded black
hole degeneracies. In section 4 we recapitulate the Rademacher
formula and Farey Tail expansion, and subsequently apply it to the
elliptic genus defined earlier. In section 5 we interpret our
result from the dual supergravity perspective and discuss its
relation to the OSV conjecture. Finally in section 6 we present
our conclusions and raise some open questions.

\vspace{.5cm}
{\it Erratum added for v2, April 22, 2008.}
The present paper relied on the properties of
Ref. \cite{Dijkgraaf:2000fq} to rewrite the ``Fareytail'' transform of
the $\mathcal{N}=(0,4)$ elliptic genus as a Poincar\'e series. Don
Zagier has noticed that some properties of the Fareytail transform,
originally assumed in Ref. \cite{Dijkgraaf:2000fq}, are erroneous. 
In this erratum, we point out where we relied on the
erroneous assumptions of \cite{Dijkgraaf:2000fq}. We refer to
Ref. \cite{Manschot:2007ha} for more details of the Fareytail transform.
 
The physical interpretation of the Poincar\'e series,
described in Section 5, remains
unchanged. Ref. \cite{Manschot:2007ha} presents a method to write the
elliptic genus  directly as a (regularized) Poincar\'e series,  
circumventing the problems of the Fareytail transform.  

The Fareytail transform is given by Eq. (4.11) in Section 4.2. This
section explains that the weight $w_\chi$ is given by
$-1-r/2$. Mimicking \cite{Dijkgraaf:2000fq}, we assumed 
that the function $\tilde \chi_\mu(\tau)$ is a proper (vector-valued)
modular form for general $w_\chi \leq 0$, including half-integer
values. This assumption is mistaken, $\tilde \chi_\mu(\tau)$
transforms only in a covariant way for integral values of $w_\chi$, or
even $r$. The appendix restores the $\bar \tau$-dependence and
applies also the Fareytail transform. In that case, the Fareytail transform
only leads to a proper modular form in case $b_2$ is even. We conclude
that the Fareytail transform of the elliptic genus of an M-theory
black hole only maps to a modular form in case the second Betti number
of the corresponding Calabi-Yau is even. The Poincar\'e series (4.12)
and (A.12) are in this case non-vanishing functions and transform
covariantly under modular transformations. The series in Eqs. (4.12)
and (A.12) in fact vanish when $r$ and/or $b_2$ are odd.

\section{Wrapped M-branes and the Near Horizon Limit}

To establish the notation, in this section we describe the BPS
bound states of wrapped M5 and M2 branes in M-theory on a
Calabi-Yau from a spacetime point of view. We will derive a
spectral flow symmetry relating states with different M2 and M5
brane charge, first from an eleven-dimensional perspective and
subsequently in terms of the effective three dimensional
supergravity that appears in the near horizon limit.

\subsection{Wrapped branes on Calabi-Yau and the spectral flow}

Consider an M5-brane (or possibly multiple M5-branes) wrapping a
4-cycle ${\cal{P}}=p^aD_a$ in the Calabi-Yau three-fold $X$. Here $\{D_a\}$ is a
basis of integral 4-cycles $H_4(X,\Z)$ in \(X\). In order for this 5-brane to be
supersymmetric, the 4-cycle ${\cal{P}}$ has to be realized as a positive
divisor. The wrapped M5-brane thus reduces to a string in the
remaining five dimensions. In addition there are five-dimensional particles
corresponding to M2-branes wrapping a two-cycle $\Sigma =q_a
\Sigma^a$, where \(\Sigma^a\) is a basis of $H_2(X,\Z)$ dual to $\{D_a\}$, i.e.
\(\Sigma^a \cap D_b = \delta_b^a \).
These particles carry charges $q_a$ under the $U(1)$
gauge fields $A_a$ which arise from the reduction of the M-theory
3-form $C$
\be
\label{CA} C = \sum_a A^a \wedge \alpha_a,
\ee
where $\alpha_a \in H^2(X,\Z)$ is a basis of harmonic 2-forms
Poincar\'e dual to the four-cycles $D_a$. Such an ensemble of
strings and particles can form a BPS bound state which leaves four
out of the eight supersymmetries unbroken.

Eventually we are interested in the
BPS states of the 4d black hole that is obtained by further
compactifying the string along an $S^1$. These states carry an
additional quantum number $q_0$ corresponding to the momentum
along the string. From the four dimensional perspective, the quantum
numbers $(p^a,q_a,q_0)$ are the numbers of D4, D2
and D0 branes in the type IIA compactification on $X$. In this
paper we will be switching back and forth between a spacetime perspective from eleven,
five, or even three dimensions, and a world volume perspective of
the M5-brane or its reduction to a world sheet.

Before going to the world volume description of the M5-brane and
its string reduction, let us describe the spectral flow of the BPS
states from a spacetime perspective. The low energy action of
M-theory contains the Chern-Simons coupling\footnote{Here and in
the following we are somewhat sloppy in writing down the proper
factors of $2\pi$ etc in effective actions. The final answer for
the spectral flow is however correctly normalized.}
\be
S_{CS}=\int C\wedge F\wedge F\;,
\ee
where $F=dC = F^a\wedge\alpha_a$ is the four-form field strength.
As a result, the M2-brane charge is defined as (here we work in
11D planck units)
\be
q_a=\int_{S^2\times S^1\times D_a }  \left( * F+C\wedge F\right)\;.
\ee
The charge thus contains a Chern-Simons type contribution
depending explicitly on the $C$-field. This term can be written as
a volume integral of $ F \wedge F$ and hence is invariant under
small gauge transformations that vanish at infinity. However, it
can still change under large gauge transformations corresponding
to shifts in $C$ by a closed and integral three form
\be
C\to C-\sum_a k^a d\sigma \wedge \alpha_a\;
\ee
with $\sigma\in[0,1]$ a coordinate for the $S^1$. This
transformation should be an exact symmetry of M-theory. The value
of the charge $q_a$, though, is not invariant but instead receives
an extra contribution proportional to the M5-brane charge $p^a$.
Namely, using
\be
\int_{D_a\times S^2\times S^1} d\sigma \wedge \alpha_b \wedge F
=d_{abc}\int_{S^2} F^c = d_{abc} p^c\;,
\ee
one finds that
\be
q_a\to q_a- d_{abc}k^bp^c \label{qflow}
\ee
under a large gauge transformation
\footnote{\(d_{abc}\) is the cubic intersection number of the Calabi-Yau \(
d_{abc} = \int_X \alpha_a \wedge \alpha_b \wedge \alpha_c.\)}.

It turns out that the $q_0$ quantum number also changes under the
spectral flow. This is most easily seen from the world volume
perspective that will be described in more details in the
following section. In fact, one has \cite{Maldacena:1997de}
\be
\label{q0flow} q_0\to q_0+k^a q_a- {1\over 2}d_{abc}k^a k^bp^c\;.
\ee
It will turn out to be convenient to introduce the symmetric
bilinear form
\be d_{ab} = -d_{abc} p^c= -\int_P \alpha_a \wedge \alpha_b, \ee
where $\alpha_a$ should be understood as the pullback of the harmonic 2-forms from the ambient Calabi-Yau \(X\) to the 4-cycle ${\cal{P}}$.

By the Lefschetz hyperplane theorem this form is non-degenerate.
In fact it has signature $(b_2-1,1)$ where $b_2=\dim H^2(X)$ is
the second Betti number of the Calabi-Yau space. Thus, for every
positive divisor ${\cal{P}}$ we obtain a natural metric $d_{ab}$
on $H^2(X,\Z)$ which turns it in to a Lorentzian lattice
$\Lambda$. From (\ref{qflow}) we see that the spectral flow
transformation amounts to shifting $q_a$ by an element of
$\Lambda$.

The Dirac quantization condition suggests that the lattice of
M2-brane charges coincides with $H^4(X,\Z)$, which can be
identified with the dual lattice $\Lambda^*$ whose bilinear form
$d^{ab}$ is given by the inverse of $d_{ab}$. However, due to the
Freed-Witten anomaly the M2-brane charge does not satisfy the
usual Dirac quantization condition, but rather \cite{Freed:1999vc,Witten:1996hc}
\be\label{qshift}
q_a \in {1\over 2} d_{abc}p^bp^c \oplus \Lambda^*.
\ee
In terms of the bilinear form \(d_{ab}\) the flow equations
(\ref{qflow}) and (\ref{q0flow}) read
\bea
 \label{qflow2}q_a &\to& q_a+ d_{ab}k^b ,\\
\label{q0flow2} q_0 &\to& q_0+k^a q_a+ {1\over 2}d_{ab}k^a k^b\;.
\eea
In this form one can see explicitly that the collection of the shifts of M2 charges
induced by all possible large gauge transformations is the Lorentzian lattice \(\Lambda\)
and that the combination
\be
\hat{q}_0=q_0-{1\over 2} d^{ab}q_a q_b
\ee
is invariant under the combined spectral flow of $q_a$ and $q_0$.
Note that due to the integrality of the symmetric form $d_{ab}$
one has $\Lambda\subset\Lambda^*$. In general, $\Lambda$ is a
proper subset of $\Lambda^*$, which means that not all charge
configurations (\(\Lambda^*\)) are related to each other by
spectral flow (\(\Lambda\)).

From the above argument we conclude that the combined spectral
flow transformations (\ref{qflow2}) and (\ref{q0flow2}) constitute
a symmetry of M-theory/string theory. This gives a very
non-trivial prediction on the BPS degeneracies that the number of
BPS states $d_{p^a} (q_a, q_0)$ should be invariant under these
transformations.

We can now compare this microscopic prediction with the macroscopic result. The leading macroscopic entropy of the 4d black hole with charges $p^a$,
$q_a$ and $q_0$ is given by \cite{Maldacena:1997de}
\be
S=2 \pi\sqrt{{\hat{q}_0 D}}
\ee
where
\be
D\equiv  {1\over 6} \int_X P \wedge P \wedge P= \frac{1}{6}\,d_{abc} \, p^a p^b p^c.
\ee
Note that $6D= - d_{ ab}\, p^a p^b$ can be interpreted as the norm
of the vector \(p^a\). The entropy formula is consistent with our
prediction that the entropy must be invariant under the spectral
flow.

Finally, we would like to point out that the spectral flow
(\ref{qflow2}) can be induced spontaneously by the nucleation of a
M5/anti-M5 brane pair with magnetic charge $k^a$, where the M5
loops through the original (circular) M5 brane before annihilating
again with the anti-M5 brane. We will make use of this comment in
the next section where this same process is translated to the near
horizon geometry.
\newline

\begin{figure}[htb!]
\centering%
\includegraphics[height=3cm]{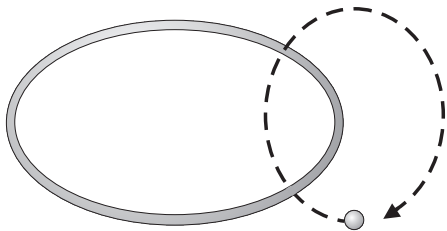}
\caption{\small{An M5 brane loops through the original (circular) M5 brane and then annihilates again with an anti-M5 brane.}}
\label{fig:Loop}
\end{figure}

\subsection{The near-horizon geometry and reduction to three
dimensions}

In the decoupling near-horizon limit, the spacetime physics can be entirely captured by the
world-volume theory of the brane . In this limit the
11-dimensional geometry becomes
\be
X \times AdS_3 \times S^2\;,
\ee
with the K\"ahler moduli $k = t^a \alpha_a
\in H^{1,1}(X)$ of the Calabi-Yau fixed by the
attractor mechanism to be proportional to the charge vector $p^a$.

More explicitly, the attractor equation reads
$$
{t^a \over V^{1/3}} = { p^a \over D^{1/3}},
$$
where $V$ denotes the volume of the Calabi-Yau
\be
V = {1\over 6} \int_X k \wedge k \wedge k = {1\over 6} d_{abc} t^a
t^b t^c\;.
\ee
Note that the volume $V$ is not fixed by the attractor equation.
Instead, the ratio $V/D$ turns out to be related to the curvature radius
$\ell$ of the $AdS_3$ and $S^2$:
$$
{V\over D} \sim {\ell_p^9 \over \ell^3}.
$$
The semi-classical limit of the M-theory corresponds to taking
$\frac{\ell}{\ell_p} \to \infty$. To keep $V/\ell_p^6$ large we
need to take $D$ very large as well. If we keep $D$ fixed, then
the CY becomes small and the 11-dimensional M-theory naturally
reduces to a five-dimensional theory. For our purpose it will be
useful to consider a further reduction along the compact $S^2$  to
a three dimensional theory on the non-compact $AdS_3$ spacetime.
In the low energy limit, this theory contains the metric and the
$U(1)$ gauge fields $A^a$ as the massless bosonic fields.

From the five-dimensional perspective, the 5-brane flux of
 M-theory background gets translated into a magnetic flux $F^a= dA^a$ of the \(U(1)\) gauge fields through the $S^2$:
$$
\int_{S^2} F^a =  p^a.
$$
The eleven-dimensional Chern-Simons term of the $C$-field can therefore
be reduced in two steps. First to five dimensions, where it takes
the form
$$
S_{CS} = \int_{AdS^3 \times S^2} d_{abc} A^a \wedge F^b \wedge
F^c,
$$
and subsequently, by integrating over the $S^2$, to three
dimensions, where it turns into the usual (Abelian)
Chern-Simons action for the gauge fields on $AdS_3$. In combination
with the standard kinetic terms, we get
$$
S = {1 \over 16\pi G_3} \int \sqrt{g}\left(R - 2\ell^{-2}\right) +
{1\over g^2} \int g_{ab} F^a \wedge \ast F^b + \int d_{ab} A^a
\wedge dA^b
$$
as the terms in the bosonic action relevant for our discussion,
where again the relative normalizations should be taken with a
grain of salt and $g_{ab}=\int_X \alpha_a \wedge \ast\alpha_b$.
The 3d Newton constant and the gauge coupling are given by
\be
{1\over G_3} \sim { V \ell^2 \over \ell_p^9} \sim {D \over \ell},\qquad\quad
{1\over g^2} \sim { V \ell^2 \over \ell_p^7} \sim {D \ell_p^2
\over \ell}.
\ee

We will end this section by some discussions about the spectral
flow in the setting of the attractor geometry. First we note that
the spectral flow argument in the previous section can be carried
to the three-dimensional setting by dimensional reduction . The
M2-brane charge $q^a$ is defined now as an integral over a circle
at spatial infinity of the $AdS_3$ as
$$
q_a = \int_{S^1} \Bigl({1\over g^2} {}^*F_a+ d_{ab}A^b\Bigr)\;.
$$
Again one easily verifies that it changes as in (\ref{qflow2}) as
a result of a large gauge transformation $A^a\to A^a + k^a
d\sigma$ in three dimensions. The charge $q_0$ is related to the
angular momentum in $AdS_3$. To understand the shift in $q_0$
under spectral flow, one has to determine the contribution to the
three-dimensional stress energy tensor due to the gauge field.

As mentioned above, the spectral flow has a nice physical
interpretation in terms of the nucleation of an M5/anti-M5 brane
pair. Let us now describe this process in the near horizon
geometry. The following argument is most easily visualized by
suppressing the (Euclidean) time direction and focusing on a
spatial section of $AdS_3$, which can be thought of as a copy of
Euclidean $AdS_2$ and hence is topologically a disk. Together with
the $S^2$ it forms a four dimensional space. First, recall that a
wrapped M5 brane appears as a string-like object in this four
dimensional space. Since an M5-brane is magnetically charged under
the five-dimensional gauge fields $A^a$, it creates a "Dirac
surface" of $A^a$.  Of course, the location of the Dirac surface
is unphysical and can be moved by a gauge transformation. Now
suppose at a certain time an M5/anti-M5-brane pair nucleates in
the center of $AdS_2$ in a way that the M5 and the anti-M5 branes
both circle the equator of the $S^2$.  Subsequently, the M5 and
the anti-M5 branes move in opposite directions on the $S^2$, say
the M5 brane to the north pole and anti-M5 to the south pole.  In
this way the M5 and anti-M5 brane pair creates a Dirac surface
that stretches between them. Eventually both branes slip off and
self-annihilate on the poles of the $S^2$. What they leave behind
now is a Dirac surface that wraps the whole $S^2$ and still sits
at the origin of the $AdS_2$. To remove it one literally has to
move it from the center and take it to the spatial infinity. Once
it crosses the boundary circle, its effect is to perform a large
gauge transformation that is determined by the charge $k^a$ of the
M5 brane of the nucleated pair. We conclude that spectral flow can
thus be induced by the nucleation of pairs of M5 and anti-M5
branes.
\newline

\begin{figure}[htb!]
\centering%
\includegraphics[height=5.5cm]{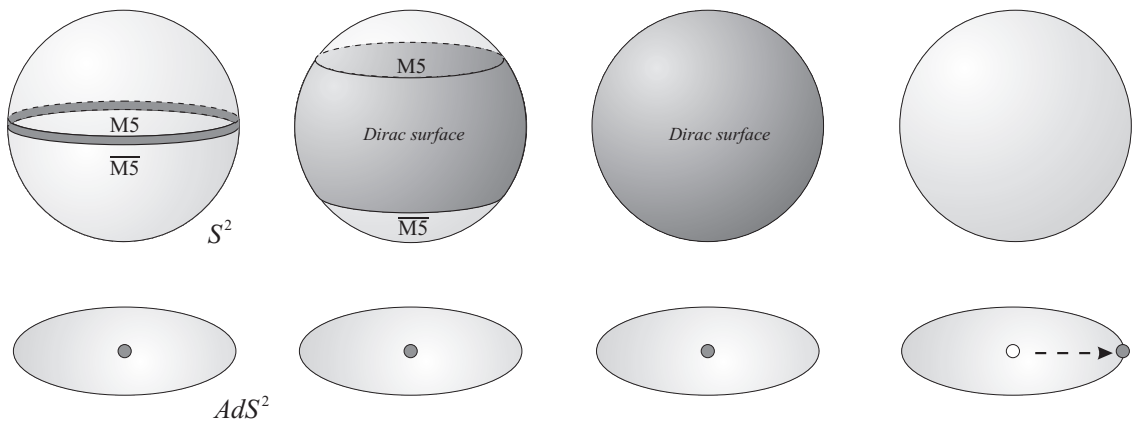}
\caption{\small{A large gauge transformation: (i) An M5-anti-M5
pair wrapping the equator of the \(S^2\)  nucleates at the center
of the \(AdS_2\). (ii) The M5 and anti-M5 begin to move in the
opposite directions in the \(S^2\), while still stay at the center
of the \(AdS_2\). (iii) A Dirac surface wrapping the whole \(S^2\)
is formed. (iv) Finally one moves the Dirac surface from the bulk
of the \(AdS_2\) towards the spatial infinity across the
boundary.}} \label{fig:nucleation}
\end{figure}

\section{The (0,4) SCFT and its Elliptic Genus}
\setcounter{equation}{0}

The existence of the bound states of M2-branes to the M5-brane can
be seen in an elegant way from the point of the view of the
five-brane world-volume theory. This world-volume theory is
a six-dimensional (0,2) superconformal field theory whose field content
includes a tensor field
with self-dual 3-form field strength $H$. The spacetime $C$ field
couples to $H$ through the term \be \label{CH} \int_Y C \wedge H
\ee where $Y = {\cal{P}} \times S^1 \times \R$ denotes the world-volume of
the five-brane.

In a bound state the M2-brane charges are
dissolved into fluxes of $H$ in the following way: the self-dual tensor $H$ that carries the
charges $q_a$ has spatial components
\be
\label{HA} H = d^{ab} \,q_a \
\alpha_b \wedge d\sigma,
\ee
with $\sigma \in [0,1]$ being the
coordinate along the $S^1$. The timelike components follow from the self-duality condition.
Combining the formulas (\ref{CA}) and (\ref{HA}), one sees that this
produces the right coupling
$$
\int_Y C \wedge H = q_a \int_\R A^a
$$
of the $U(1)$ gauge fields \(A^a\) to the charges $q_a$.

\subsection{The (0,4) superconformal field theory}

When we take the scale of the Calabi-Yau to be much smaller than
the radius of the M-theory circle, the M5 world-volume theory
naturally gets reduced along the
4-cycle ${\cal{P}}$ to a two-dimensional conformal field theory with
$(0,4)$ supersymmetry.
As usual the superconformal symmetries are identified
with the supersymmetric isometries of the $AdS_3 \times S^2$
manifold. This algebra contains the following generators that will be important for us: the
right-moving stress-energy tensor $\overline{T}$, supercurrents
$\overline{G}_\alpha^\pm$ and  $SU(2)$ R-symmetry currents
$\overline{J}_{\alpha\beta}$. In particular the $SU(2)$ R-symmetry
corresponds to the rotations of the $S^2$ factor.

The self-dual tensor field \(H\) gives rise to a collection of chiral scalar
fields $\varphi^a$ through an expansion of  $H$ in harmonic forms.
To be more specific, if we write its spatial components as
\be
H = \imath^{\ast}\alpha_a \wedge \partial_\sigma \varphi^a
d\sigma\;,
\ee
the self-duality condition\footnote{In the presence of a
non-trivial background field $C$ the self-duality equations are
modified. This shows up in the CFT as Narain moduli for the chiral
bosons. The dependence of the elliptic genus on the Narain moduli
is easy to work out, they only appear in the theta functions
(\ref{j2}) and in particular the entropy does not depend on them.}
automatically determines the temporal components. Here
$\imath^{\ast} \alpha_a\in H^2({\cal P},\mathbb Z)$ is the
pull-back of the harmonic two-form $\alpha_a$ under the inclusion
map $\imath:{\cal P} \hookrightarrow X$. The two-forms in
$H^2({\cal P},\mathbb Z)$ that are not in the image of
$\imath^{\ast}$ are not conserved due to M2-brane instantons and
do not give rise to chiral scalar fields
\cite{Maldacena:1997de,Minasian:1999qn}. Altogether one concludes
that the self-dual and anti-self-dual two-forms on $\cal P$ that
do extend to non-trivial two-forms on $X$ give rise to the left-
and right-moving scalars in the CFT respectively. Using the fact
that the K\"ahler form $k$ is the only self-dual two form among
all the harmonic 2-forms $\imath^{\ast} \alpha_a \in H^2({\cal P};
{{\mathbb Z}})$, one finds one right-moving and $(b_2-1)$
left-moving bosons from the reduction of the self-dual tensor
field \(H\) of the five-brane world volume theory.

In addition there are three non-chiral scalar fields $x^i$ that
describe the motion of the five-brane transverse to the
Calabi-Yau. The four right-moving scalars assemble themselves into
a $(0,4)$ supermultiplet together with the four right-moving
Goldstinos $\overline{\psi}_\alpha^\pm$ arising from the broken
supersymmetry. This supermultiplet of free fields, which we will refer to as the
"universal" supermultiplet \cite{Minasian:1999qn}, represents only
a small part of the full conformal field theory, since it only
gives a fixed (\({\cal{P}}\) independent) contribution to the central charge. The bulk of the
degrees of freedom are contained in a separate sector which can be
interpreted as describing the deformation of the five-brane inside the
Calabi-Yau. In the semi-classical limit this CFT will take the form of
a heterotic sigma-model on the relevant moduli space. Especially, the
left-moving and right-moving central charges are \cite{Maldacena:1997de}
\be
c_L = 6 D + c_2 \cdot P\;,\qquad c_R = 6 D + {1\over 2} c_2 \cdot P\;.\ee

For the following discussion it will be useful to spell out in
more details the definition of the left-moving and right-moving
components of the bosons \(\varphi^a\) . The fact that the K\"ahler form takes the form
$k \sim [{\cal{P}}] = p^a\alpha_a$ implies that there is  only a single right-moving component
corresponding to  the component of $\varphi^a$ in the direction of $p^a$. Let us therefore define
the left- and right-moving projectors
\be
P_{R,b}{}^a=- \frac{d_{bc}\,p^ap^c}{6D} = \delta_b{}^a -
P_{L,b}{}^a\;.
\ee
One thus finds
that the left-movers obey
\be
P_{R,a}{}^b (\del  \varphi^a_L) = 0 \Leftrightarrow
\del \varphi^a_L = P_{L,b}{}^a (\del \varphi^b)
\ee
while the right-mover is defined through
\be
 \bar{\del} \varphi^a_R = P_{R,b}{}^a (\bar
{\del} \varphi^b)\;.
\ee
The charges $q_a$ correspond to the momenta of the scalars
$\varphi^a$, thus a membrane of charge $q_a$ is represented in the CFT
by an insertion of the vertex operator
\be
V_q = \exp i\left(q_{L,a}\varphi_L^a+ q_{R,a} \varphi_R^a
\right),
\ee
where $q_L$ and $q_R$ are defined in terms of the (half) integral
charges $q_a$ as
\be
q_{L,a} =  P_{L,a}{}^b q_b, \qquad
q_{R,a} =
P_{R,a}{}^b q_b\;.
\ee
Furthermore one has \cite{Maldacena:1997de}
\be
q_L^2 - q_R^2 = d^{ab} q_a q_b.
\ee

Later we will be interested in the supersymmetric
ground states in the charge sectors labelled by $q_a$. A special
role in this analysis will be played by the right-moving universal
multiplet. It will be convenient to introduce the
right-moving field by $\overline{\varphi}$ defined by
\be
\varphi_{R}^a = p^a \overline{\varphi}.
\ee
It is normalized so that its operator product expansion reads
$$
\delbar\overline{\varphi}(\zbar) \cdot
\delbar\overline{\varphi}(\wbar) \sim - {1\over 6D} {1\over (\zbar
- \wbar)^2}
$$
We also normalize the Goldstinos
$\overline{\psi}^\pm_\alpha$ which sit in the same supermultiplet
in a similar fashion, so that the following identities holds \bea
\epsilon^{\alpha\beta}\left\{ \overline{\psi}^+_{\alpha,0},\overline{\psi}^-_{\beta,0}
\right\} 
 & \! = \! & {1\over 3 D}
, \nonumber \\
\epsilon^{\alpha\beta}\left\{ \overline{G}^+_{\alpha,0},\overline{\psi}^-_{\beta,0}
\right\} 
 & \! = \! &
\frac{p^a q_a}{3D}\, \label{com} \\
\epsilon^{\alpha\beta}\left\{ \overline{G}^+_{\alpha,0},\overline{G}^-_{\beta,0}
\right\} 
 & \! = \! &  4 \left( \overline{L}_0-{c_R\over 24}\right) . \nonumber
\eea

The spectral flow relations (\ref{qflow2}) (\ref{q0flow2}) are
implemented in the (0,4) CFT as a symmetry of the superconformal
algebra. It is given by
\begin{eqnarray}
&L_m&\to L_m+k_L^a Q_{aL,m} - \frac{1}{2} \,d_{abc}k_L^a k_L^b
p^c\delta_{m,0}
\nonumber \\
&Q_{aL,m}&\to Q_{aL,m} - d_{abc}k_L^b p^c\delta_{m,0}\;,
\end{eqnarray}
where \(Q_{aL,m}\) are the modes of the \((b_2-1)\) left-moving
bosons, and with similar expressions for $\bar{L}_m$ and
$Q_{aR,m}$. Note that these transformations leave
$L_0-\frac{1}{2}Q^2_{L}$ and $\bar{L}_0-\frac{1}{2}Q^2_{R}$
invariant.

\subsection{A generalized elliptic genus}

The generating function of BPS bound states for a fixed M5-brane
charge $p^a$ can be identified with a generalized elliptic genus
of the CFT (see also
\cite{Denef-Moore,Gaiotto:2006,Kraus:2006nb}). More precisely, we
want to compute the partition function
\be \label{part}
Z'_{p^a}(\tau,\taubar;y^a)= \Tr\left[ F^2 (-1)^F e^{\pi ip^aq_a}
e^{2\pi i \tau \left( L_0-\frac{c_L}{24}\right)}e^{-2\pi i \taubar
\left( \bar{L}_0-\frac{c_R}{24}\right)} e^{2\pi i y^a q_a} \right]
\;,
\ee
where $F$ denotes the fermion number $F=2\bar{J}^3_0$, with
$\bar{J}^3_0$ one of the generators of $SU(2)_R$. The appearance
of the extra phase $e^{\pi i p^a q_a}$ is a bit subtle and will be
explained below. Notice that we could also have chosen to absorb
this phase in the definition of fermion number, but instead we
will keep it explicit in what follows. The reason that we have to
insert $F^2$ inside the trace is to absorb the four fermion
zero-modes and therefore obtain a non-zero expression
\cite{Maldacena:1999bp}. The trace with an insertion of $F^k$ with
$k=0,1$ vanishes identically.

Only short multiplets, namely those states that are annihilated by four
combinations of world-sheet supercharges
$\overline{G}^\pm_{\alpha,0}$ shifted by an appropriate multiple
of the fermion zero-mode $\overline{\psi}^\pm_{\alpha,0}$, contribute to the partition
function (\ref{part}). In the
sector with \(U(1)\) charges $q_a$, these states satisfy
\be \label{susy1}
\left(\overline{G}^\pm_{\alpha,0} - p^a q_a
\overline{\psi}^\pm_{\alpha,0}\right) |BPS\rangle = 0.
\ee

Using the fact that the norm of these
states is positive, one derives that
\be\label{barL}
\epsilon^{\alpha\beta}\left\{\left(\overline{G}^+_{\alpha,0} - p^a q_a
\overline{\psi}^+_{\alpha,0}\right), \left(\overline{G}^-_{\beta,0} - p^a q_a
\overline{\psi}^-_{\beta,0}\right)\right\} |BPS\rangle  =  4\,\,
\left(\overline{L}_0-{1\over 2}q_R^2-{c_R\over 24}\right)
|BPS\rangle = 0.
\ee
where
\be\label{rightmovingmetric}
q_R^2 \equiv   - d^{ab}\,q_{R,a}q_{R,b} ={(p^aq_a)^2\over 6D}
\;.
\ee
It is straightforward to show that once we include the
three-momentum $\vec{p}$ of the non-compact scalars of the
universal supermultiplet, the BPS condition becomes $$ \bar{L}_0 =
\frac{1}{2} q_R^2 + \frac{1}{2} \vec{p}^2 + \frac{c_R}{24}. $$ In
each BPS multiplet, there are four linear combinations of
$\bar{G}^{\pm}_{\alpha,0}$ and $\bar{\psi}^{\pm}_{\alpha,0}$ that
 act non-trivially and produce a short
multiplet of 4 BPS states. A direct consequence of this is that
the partition function depends in a specific way on $\taubar$. \
In the sector with $\vec{p}=0$ this dependency is captured by the
``heat equation''
\be\label{heateq}
\left\lbrack \partial_{\taubar} + {1\over 24\pi i D }
\left(p^a{\partial_{y^a}}\right)^2\right\rbrack
Z'_{p^a}(\tau,\taubar,y) =0,
\ee
which means that the dependence on $\taubar$ is completely under
control. It turns out that it only appears in the elliptic genus
through certain theta functions, which is discussed in detail in
appendix~A.

Eventually our aim is to show that the partition function has an
asymptotic expansion in terms of semi-classical saddle-points of
the three-dimensional supergravity theory. A crucial property of
$Z'_{p^a}(\tau,\taubar;y^a)$ for this is that it is a modular
form. To be precise, it is a modular form of weight $(0,2)$. To
show this, let us introduce a generalized partition function \be
\label{partW} W_{p^a}(\tau,\taubar;y^a, z)= \Tr\left[ e^{2\pi i z
F} e^{\pi i p \cdot q} e^{2\pi i \tau
    \left( L_0-\frac{c_L}{24}\right)}e^{-2\pi i \taubar \left(
    \bar{L}_0-\frac{c_R}{24}\right)} e^{2\pi i y^a q_a} \right] \ee

\noindent so that
$$
Z'_{p^a}(\tau,\taubar;y^a)=-{1\over 4\pi^2} \partial_z^2
W_{p^a}(\tau,\taubar;y^a, z)|_{z=1/2}\;.
$$
The partition function $W_{p^a}(\tau,\taubar;y^a, z)$ can be
represented as some kind of ``functional integral'' over all
degrees of freedom of the (0,4) CFT on a torus with modular
parameter $\tau$ and certain Wilson lines parametrized by $y^a$
and $z$. As such it should be independent of the choice of cycles
on the torus, and hence be invariant under the modular
transformations
$$
\tau\to {a\tau+b\over c\tau+d},\qquad \taubar\to {a\taubar+b\over
c\taubar+d}, \qquad y_L^a\to \frac{y_L^a}{c\tau+d}, \qquad
y_R^a\to \frac{y_R^a}{c\taubar+d}, \qquad
z\to\frac{z}{c\taubar+d}\,,
$$
where \(y_L, y_R\) are the projected potentials \(y_L^a =
P_{L,b}{}^a\,y^b\), \(y_R^a = P_{R,b}{}^a\,y^b\). \noindent This
proves that $Z'_{p^a}(\tau,\taubar;y^a)$ has weight $(0,2)$.

As mentioned above, the partition function
$Z'_{p^a}(\tau,\taubar;y^a, z)$ contains a continuous degeneracy
in the BPS states due to the zeromodes in the $\mathbb R^3$ part
of the $S^1 \times \mathbb R^3$, $\mathcal{N}=(0,4)$ "universal"
multiplet. We wish to extract this degeneracy, and this can be
done by defining
\be \label{mod5}
Z'_{p^a}(\tau,\bar{\tau},y^a,z) = \int d^3 p (e^{2\pi i \tau} e^{
-2\pi i \bar{\tau}} )^{\frac{1}{2} \vec{p}^2}
Z_{p^a}(\tau,\bar{\tau},y^a,z)
\ee
where $Z_{p^a}$ is the index without zero modes
\footnote{Here the \(p\)'s refer to the momenta in
$\mathbb R^3$, not to be confused with the M5 brane charges \(p^a\).}. The integral is
proportional to ${\rm Im}(\tau)^{-3/2}$ which has weight
$\left(\frac{3}{2},\frac{3}{2}\right)$. So we establish that
$Z_{p^a}$ has weight $\left(-\frac{3}{2},\frac{1}{2}\right)$.

From (\ref{heateq}), or equivalently, (\ref{barL}), we see that we
can always reconstruct the \(\bar{\tau}\)-dependence of the
elliptic genus from the \(y\) dependence. Physically this
corresponds to the fact that for BPS states, \(\bar{L}_0\) is
determined uniquely by the \(U(1)\) charges \(q_a\). Therefore,
for clarity we will set \(\tau =\bar{\tau}\) in the following part
of the paper without losing any generality, as the full
computation with \(\taubar\)-dependence incorporated will be given
in the appendix.


For the purpose of extracting the  degeneracy, we would like to write $Z_{p^a}$ in a
Fourier expansion as
\be \label{notaubar}
Z_{p^a}(\tau,y)=\sum_{q_0; \;q_a\in
  \Lambda^*+d_{ab}p^a/2} d(q_0,q)  \,e^{2\pi i\tau q_0}\,e^{2\pi i y^a q_a}\;.
\ee
In more detail, for supersymmetric states one has\footnote{The
appearance of the extra \(\frac{c_R}{24} \) is due to the fact
that we are computing the elliptic genus in the R sector of the
CFT, while this term will be absent on the right-moving
supersymmetric side in the NS sector, where the supergravity
computation is carried out.}
\bea
L_0 &=& q_0^{\rm ind} + \frac{d^{ab}q_{L,a}q_{L,b} }{2} \nonumber \\
\bar{L}_0 -\frac{c_R}{24} &=&   -\frac{d^{ab}q_{R,a}q_{R,b} }{2} \\
\Rightarrow q_0 &=& ( L_0  -\frac{c_L}{24} )- (\bar{L}_0
-\frac{c_R}{24} ) = q_0^{\rm ind} + \frac{d^{ab}q_aq_b}{2}
-\frac{c_L}{24}\;, \nonumber
\eea
where $q_0^{\rm ind}$ is the induced D0-brane charge. This need
not be an integer but supersymmetry requires $q_0^{\rm ind}\geq
0$. As argued in section 3.1, large gauge transformations shift
the charges as (\ref{qflow2}), (\ref{q0flow2}). This should be a
symmetry of M-theory and leave the degeneracies of the BPS states
invariant. In the language of the dual CFT, these large gauge
transformations correspond to spectral flows which flow the
\(U(1)\) charges as (\ref{qflow2}) while keeping
\be \label{qhat}
\qhat \equiv q_0  -\frac{1}{2} d^{ab} q_a q_b = q_0^{\rm ind}
-\frac{c_L}{24}
\ee
invariant. This is a consequence of the fact that the spectral
flow operation, while shifting the zero-modes of the \(U(1)\)
currents and thus shifting their contribution to the momentum
along the \(S^1\), permutes with all the other operators in the
CFT and thus leaves the rest of the \(S^1\) momentum unchanged.

An important subtlety is that the appearance of the extra phase
$e^{\pi i p\cdot q}$ in (\ref{part}), on top of $(-1)^F$. This
additional phase implies that under spectral flow with charge
vector $k^a$, states in the Ramond sector pick up an extra phase
$(-1)^{p\cdot k}$. This is related to the shift $p_a/2=
d_{ab}p^b/2 $  in the charges $q_a$  due to the Freed-Witten
anomaly. This can be seen, for example, from the operator algebra.
Consider the vertex operators $V_{k^a}(z)$ acting on the state
$\left.| q\right>$  with $q_a\in\Lambda^*+ p_a/2$, one has
\begin{equation} \label{aux9}
V_{k^a}(z) \left.| q\right>=z^{k\cdot q}\left.|q+k\right>\;.
\end{equation}
The OPE will pick up a phase $\exp\left(2\pi i k\cdot q
\right)=(-1)^{k\cdot p}$ when $z$ circles around the origin.
Locality of the OPE requires projection onto the states with even
$k\cdot p$, which explains why the elliptic genus needs to contain
a factor $(-1)^{p\cdot k}$ for it to be modular invariant. For
convenience we include however a factor $e^{\pi i p\cdot q}=e^{\pi
i p\cdot\mu}e^{\pi i p\cdot k}$ in our definition of the elliptic
genus.  The term $p\cdot\mu$ could be interpreted as an additional
overall phase or as a fractional contribution to the fermion
number.


Substitution of the transformations (\ref{qflow2}) and
(\ref{q0flow2}) in (\ref{notaubar}) shows directly the transformation
property of the partition function under spectral flow to be
\be
Z_{p^a}(\tau,y)=e^{\pi i p\cdot k}e^{\pi i \tau k^2 +2\pi i y\cdot k }Z_{p^a}(\tau,y+k\tau),
\ee

\noindent where we have taken the fermion number as well into
account. Spectral flow therefore predicts\footnote{To keep the transparency of the equations, we leave out the indices in
various places in this section. But it should be understood that all the indices are lowered, raised,
and contracted by using \(d_{ab}\) and \(d^{ab}\).}
\be\label{cflow}
d(q_0,q) = e^{\pi i p\cdot k}d(q_0+k\cdot q + \frac{k^2}{2}, q+ k ) \;.
\ee
From this invariance property and the shift of the charge lattice due to the anomaly
(\ref{qshift}), it turns out to be convenient to decompose the \(U(1)\) charges as
\be
q= \mu + k \;\;\;\;;\;\;\;\; \mu \in \Lambda^{\ast}/\Lambda+\frac{p}{2}\;\;\;\;;\;\;\;\; k \in \Lambda\;.
\ee
In this way we split the sum over the charges into a sum over \(\mu\) in the "fundamental domain" \( \Lambda^{\ast}/\Lambda+\frac{p}{2}\) and a sum over the different possible spectral flows \(\Lambda\).
Exploiting (\ref{cflow}), we now rewrite the elliptic genus as
\be
Z_{p^a}(\tau,y)=\sum_{\mu\in \Lambda^{\ast}/\Lambda+\frac{p}{2}} \,\sum_{k \in \Lambda}
\,\sum_{ -\frac{c_L}{24} \leq \qhat} e^{\pi i p\cdot \mu} e^{\pi i p\cdot k}
d_{\mu}(\qhat) \, e^{2\pi i \tau \left(\qhat+ \frac{(\mu+k)^2}{2}\right) }\;
e^{2\pi i y\cdot(\mu +  k)}\;,
\ee
where \(d_{\mu}(\qhat) = (-1)^{k\cdot p}
d(\qhat+\frac{(\mu+k)^2}{2},\mu+k\)). Putting all the factors
together, we conclude that our generalized elliptic genus can be
decomposed into purely holomorphic modular forms which contain all
the information about the degeneracies and theta-functions (this
is in fact also a general property of weak Jacobi forms, see e.g.
\cite{zagier})
\be\label{decomposetheta}
Z_{p^a}(\tau,y)= \sum_{\mu\in \Lambda^{\ast}/\Lambda+\frac{p}{2} }\chi_{\mu}(\tau)\; \theta_{\mu}(\tau,y)\;,
\ee
where
\bea
\chi_{\mu}(\tau)&=& \sum_{ -\frac{c_L}{24}\leq \qhat}d_{\mu}(\qhat)\;e^{2\pi i \tau \qhat}\\
\theta_{\mu}(\tau,y)&=& \sum_{k\in \Lambda} e^{\pi i p\cdot \mu} e^{\pi i p\cdot k} e^{2\pi i \tau \frac{(\mu+k)^2}{2}+2\pi i y\cdot (\mu + k)}\;.
\eea
Especially, the degeneracies in a sector with given $q_a$ are contained purely in
$\chi_{\mu}(\tau)$, as there is only one term in the theta
functions that contributes for each value of $q_a$. And the necessity of the factor
$\left(-1\right)^{p\cdot (\mu+k)}$ can also be seen from the fact that it ensures that the theta functions
transform into themselves under modular transformations.

Since the lattice $\Lambda$ is Lorentzian, the holomorphic theta functions are only defined formally, since the sum is not convergent. To obtain a convergent
expression one has to restore the $\taubar$-dependence. We will discuss this in more detail in the appendix, but for completeness we give here the
resulting  $\taubar$-dependent partition function. It takes the form
\be
Z_{p^a}(\tau,\bar{\tau},y_L,y_R) =
\sum_{\mu\in\Lambda^{\ast}/\Lambda+\frac{p}{2}}
 \chi_{\mu} (\tau) \theta_{\mu} (\tau,\bar{\tau},y_L,y_R),
 \ee
where $\chi_\mu$ is defined as above. Hence, the only difference is that the theta-function is replaced by
a so-called Siegel-Narain theta function:
\be \label{j2}
\theta_\mu(\tau,\taubar, y_L, y_R) = \sum_{k\in \Lambda}
 e^{\pi i p\cdot \mu} e^{\pi i p\cdot k} e^{2 \pi i \tau\frac{(\mu_L+k_L)^2}{2} } \,e^{-2 \pi i \taubar\frac{(\mu_R+k_R)^2}{2} }
e^{2\pi i y \cdot (\mu+k)}
 \;,
\ee
which is convergent due to the presence of \(\taubar\)-dependence. A
discussion of the modular properties of these theta functions and a
further explanation of these expressions
can be found in the appendix and in e.g. \cite{borcherds-1998-132}.

\section{The Farey Tail Expansion of the Elliptic Genus}
\setcounter{equation}{0}

Now that we have obtained an expression for the elliptic genus, as the following step  we would like to rewrite it
in the form of an asymptotic expansion that is suitable for a semi-classical interpretation.
The main tool we will use for this purpose is the so-called Rademacher formula.
This formula can be applied to practically any modular form and is rather insensitive to
most of the details of the system. Therefore, to exhibit the main idea we will in this first subsection present the general arguments that lead to the
Rademacher formula. Subsequently we will generalize it so that it can be applied to the case of interest.

\subsection{The Rademacher formula}

An essential ingredient of the Farey tail is the Rademacher formula, which is an
asymptotic expansion given by a sum over images of the "polar part" of the partition function under the modular group, which we will define shortly.
The name "Farey tail" comes from the fact that the terms in the expansion corresponding
to different images under the modular group are labelled by a Farey series, which is a sequence of rational
numbers ordered by the size of the denominator. One term is the
leading term of the sequence whereas the others form a ``tail'' of
subleading corrections.

To illustrate the basic idea of the Rademacher formula, let us
consider, as the first case, a modular form $Z(\tau)$ of weight $w$
with Laurent expansion
\be\label{Zexpansion}
Z(\tau) = \sum_{n\geq 0} d(n) e^{2\pi i(n-{c\over 24})\tau}.
\ee
Here $d(n)$ are integer degeneracies.  The fact that it has weight \(w\) means that
under modular
transformations
it transforms as
\be
Z(\tau)=Z\left(a\tau+b\over
c\tau+d\right){\left({c\tau+d}\right)^{-w}},\qquad \left(
 \begin{array}{cc} a& b\\ c&d \end{array}\right)\in SL(2,\Z).
\ee
The modular form $Z(\tau)$, when expressed in terms of $q= \exp 2\pi i\tau$,
has poles at the `origin' $q=0$ and its images under the modular
group.

The Rademacher formula is obtained by first truncating the sum
over $n$ to include only the polar terms, that is, those terms with \(n < \frac{c}{24}\),
and subsequently summing over all the images under the modular group. This procedure
reproduces the original modular form $Z(\tau)$, since by
construction one obtains a modular form with the same weight and
poles at exactly the same location. The Rademacher formula thus
reads
\be
Z(\tau)= \sum_{\gamma\in
\Gamma/\Gamma_\infty}Z^-\left({a\tau+b\over c\tau+d}\right)
\left({c\tau+d}\right)^{-w}
\ee
with
\be
Z^-(\tau)= \sum_{0\leq n<{c\over 24}}d(n)  e^{2\pi
i\tau(n\!-\!{c\over 24})}\;.
\ee
Here $\gamma = \left(\begin{array}{cc} a& b\\ c&d \end{array}\right)$
and $\Gamma_\infty$ denotes the modular subgroup that keeps
$\tau=i\infty$ fixed and is generated by $\tau\to\tau+1$. To
ensure that the sum over the modular group is convergent, the
modular weight $w$ of $Z(\tau)$ should be positive.

The Rademacher formula can be used to express the degeneracies $d(n)$
of arbitrary $n$ in terms of those corresponding to the polar part,
and this leads to an exact version of the Cardy formula. In integral form it reads
\be
d(m) =\sum_{0\leq n<{c\over 24}} d(n) \sum_{\gamma\in
\Gamma_\infty\backslash \Gamma'/\Gamma_\infty} K_\gamma(n;m)
\ee
with
\be
K_\gamma(n;m)= {1\over 2\pi}  \int_{-\infty}^\infty \! d\tau\,
e^{2\pi i\left( \frac{a\tau+b}{c\tau+d}(n-{c\over 24})-\tau (m-{c\over
24})\right)}\left({c\tau+d}\right)^{-w}\;,
\ee
where the prime of \(\Gamma' \) indicates that we leave out the identity element
of the modular group. Here we have performed part of the summation over
the modular group to extend the range of integration over $\tau$
to the full real axis. The resulting integral can be expressed in
terms of generalized Bessel functions. For more details, see
\cite{Dijkgraaf:2000fq}.

Both for the D1-D5 system as well as for the present case of
attractor black holes, one needs a generalization of the
Rademacher formula, namely one that applies to vector-valued
modular forms (or characters) whose transformation properties involve a
non-trivial representation of the modular group. In both of these
cases one encounters characters $\chi_\mu(\tau)$ with an expansion
of the form
\be
\chi_\mu(\tau)=\sum_n d_\mu(n)e^{2\pi i\tau(n+\Delta_\mu-{c\over
24})}
\ee
in terms of integral coefficients $d_\mu(n)$. The modular
transformation rules take the form
\be
\chi_\mu(\tau) =
M(\gamma)_\mu{}^\nu\chi_\nu\left(a\tau+b\over c\tau+d\right)
 \left({c\tau+d} \right)^{-w_{\chi}}
 \ee
where the matrices $M(\gamma)$ form a faithful representation of
the modular group and \(w_{\chi}\) is the weight of the character \(\chi_{\mu}\). In this case the conformal weights $\Delta_\mu$
are not necessarily integers anymore. By applying a similar
reasoning as above, one obtains a generalized Rademacher formula
involving the matrices $M(\gamma)$. In full glory it reads
\be
\chi_\mu(\tau)= \sum_{\gamma\in
\Gamma/\Gamma_{\infty}}M(\gamma)_\mu{}^\nu
\sum_{n+\Delta_\nu<{c\over 24}}  d_\nu (n)
 e^{2\pi i\frac{a\tau+b}{c\tau+d}(n+\Delta_\nu-{c\over 24})} \left({c\tau+d} \right)^{-w_{\chi}}
\ee
We eventually want to apply the formula to an elliptic genus
$Z(\tau,y)$ that in addition to the modular parameter $\tau$
depends on one or more potentials (or wilson lines) $y^A$,
$A=1,..,r$. The crucial property of the elliptic genus that
allows one to apply the Rademacher formula, is that it  factorizes
in terms of theta functions as
$$
Z(\tau,y)=\sum_\mu \chi_\mu(\tau)\theta_{\mu}(\tau,y).
$$
The rank of the theta function equals the number $r$ of the potentials
$y^A$. A necessary and sufficient condition for $Z(\tau,y)$ to
factorize in this way is that it obeys the following "spectral
flow" property
$$
Z(\tau,y+n\tau+m)=e^{ -i\pi \tau n^2 \tau-2\pi i n y}Z(\tau,y)
$$
and it transforms under $\tau\to \frac{a\tau+b}{c\tau+d}$ and $y\to \frac{y}{c\tau+d}$ as
$$
Z\left(\frac{a\tau+b}{c\tau+d},\frac{y}{c\tau+d}\right)= e^{i\pi
c y^2\over c\tau+d} Z(\tau,y)(c\tau+d)^{w}
$$
These transformation properties imply that $Z(\tau,y)$ is a weak
Jacobi-form of weight $w$. Given the fact
that the theta functions $\theta_{\mu}(\tau,y)$ form a
representation of the modular group and have modular weight equal
to ${1\over 2}r$, it follows that the characters $\chi_\mu(\tau)$
transform in the conjugate (=inverse) representation and have a
modular weight
$$
w_\chi=w-{1\over 2}r\;.
$$
This fact can now be used to write the
Rademacher formula for the full partition function as
\be\label{rademacher}
Z(\tau,y)=\sum_{\gamma\in
\Gamma/\Gamma_{\infty}}e^{-\pi i \frac{c\,y^2}{c\tau+d}} \chi_\mu^-\left({a\tau+b\over c\tau+d}\right)\theta_\mu\left({a\tau+b\over c\tau+d},{y\over c\tau+d}\right)\left({c\tau+d} \right)^{-w}\,
\ee

The Rademacher expansion is primarily acting on the vector valued
modular forms $\chi_\mu(\tau)$ when applied to a weak Jacobi form. In case $w_\chi<0$, first a so called
``Farey tail transform''\cite{Dijkgraaf:2000fq} has to be done to get
a convergent answer, given by
\be \label{eq:fareytrans}
\tilde \chi_\mu(\tau)=(q\frac{\partial}{\partial q})^{1-w_\chi}\chi_\mu(\tau)\;.
\ee
The transformed function has weight $\tilde w_\chi=2-w_\chi$, which shows that the Rademacher
expansion can be applied to this modular form in case $w_\chi<0$.

\subsection{Application to the (0,4) elliptic genus}

We now like to apply the Rademacher formula to the elliptic genus
of the (0,4) SCFT and obtain an exact rewriting of the generating
function \(Z_{p^a}(\tau,y)\) of the BPS degeneracies \(d_\mu(n)\).
As stated above, one needs $\chi_\mu(\tau)$ to have a weight
$w_\chi>0$ for the expansion to converge \cite{Dijkgraaf:2000fq}.
\(Z_{p^a}(\tau,y)\) has a negative weight $-{3 \over 2}+{1\over 2}
=-1$. From which we deduce that $\chi_\mu(\tau)$ has weight
$-1-\frac{1}{2}r$, which is manifestly negative. In order to get a
meaningful and convergent Rademacher expansion, we apply the Farey
tail transformation eq. (\ref{eq:fareytrans}). This gives a
transformed modular form with weight $\tilde w_\chi=3+{1\over
2}r$. Combining it again with the theta functions gives a weak
Jacobi form $\tilde Z_{p^a}(\tau,y)$ of weight $w=3+r$.

The full answer is of course similar to (\ref{rademacher}), and
reads
\be \label{rad2}
\tilde{Z}_{p^a}(\tau,y) =
 \sum_{ \mu\in \Lambda^{\ast}/\Lambda+\frac{p}{2} }
 \,\sum_{ \gamma\in\Gamma/\Gamma_{\infty} }  e^{-\pi i \frac{c\,y^2}{c\tau+d}}
 \tilde{\chi}_\mu^- \left( \frac{a\tau+b}{c\tau+d}\right)
 \theta_\mu\left(\frac{a\tau+b}{c\tau+d},\frac{y}{c\tau+d}\right)\,\left({c\tau+d} \right)^{-(3+r)}\,
\;,
\ee
where
\be
\tilde{\chi}_\mu^-(\tau)= \sum_{ -\frac{c_L}{24} \leq \qhat < 0}
\tilde{d}_{\mu}(\qhat)\;e^{2\pi i \tau \qhat}, \qquad \tilde
d_\mu(\hat q_0)= (\hat q_0)^{2+{1\over 2}r} d_\mu(\hat q_0),
\ee
now contains only the "polar part". This is our Attractor Farey
Tail: the Rademacher expansion for the elliptic genus which is the
generating function of \({\cal{N}}=2\) D4-D2-D0 BPS black hole
degeneracies. Notice that both the Farey tail transform as well as
the Rademacher expansion commute with putting $\tau=\bar{\tau}$.
In other words, if we would have kept the $\bar{\tau}$ dependence
all along and would only have put $\tau=\bar{\tau}$ at the end we
still would have ended up with the same result (\ref{rad2}).

\section{Spacetime Interpretation of the Attractor Farey Tail}
\setcounter{equation}{0}

So far the Rademacher formula appears to be just a mathematical
result. What makes it interesting is that it has a very natural
interpretation from the point of view of a dual gravitational
theory. In this section we discuss the interpretation of the Farey tail expansion first in
terms of the effective supergravity action, and
subsequently from an M-theory/string theory perspective.
We will first discuss the gravitational interpretation of the
general formula presented in section 4.1, and then turn to the present
case of the attractor black holes.

\subsection{Gravitational interpretation of the Rademacher formula}

Microscopic systems described by a 2d CFT have a dual description
in terms of a string- or M-theory on a space that contains $AdS_3$
as the non-compact directions.  This is because $AdS_3$ is the
unique space whose isometry group is identical to the 2d conformal
group. The miracle of AdS/CFT is that the dual theory contains
gravity, which suggests that the partition function of the 2d CFT
somehow must have an interpretation as a sum over geometries. The
full dual theory is defined on a space that is 10- or
11-dimensional, but except for the three directions of $AdS_3$
these dimensions  are all compact. Hence, by performing a
dimensional reduction along the compact directions we find that
the dual theory can be represented as a (super-)gravity theory on
$AdS_3$. The effective action therefore contains the  Einstein
action for the 3d metric
$$
S_E={1\over 16\pi G_3} \int_{AdS}\!\!\! \sqrt{g} (R-2\ell^{-2})+{1\over 8\pi
G_3} \int_{\del (AdS)}\!\!\! \sqrt{h} (K-\ell^{-1})
$$
where we have included the Gibbons-Hawking boundary term. Here
$\ell$ represents the AdS-radius. According to the AdS/CFT
dictionary, the 3d Newton constant $G_3$ is related to the central
charge $c$ of the CFT by \be {3\ell\over 2G_3}= c\;. \label{Gc} \ee
The dictionary also states that the partition function $Z(\tau)$
of the CFT is equal to that of the dual gravitational theory on (a quotient of)
$AdS_3$, whose boundary geometry coincides with the 2d torus on
which the CFT is defined. The shape of the torus is kept fixed and
 parametrized by the modular parameter $\tau$.

The rules of quantum gravity tell us to sum over all possible
geometries with the same asymptotic boundary conditions. For the
case at hand, this means that we have to sum over all possible three dimensional
geometries with the torus as the asymptotic boundary. Semi-classically,
these geometries satisfy the equations of the motion of the
supergravity theory, and hence are locally $AdS_3$. There
indeed exists an Euclidean three geometry with constant curvature
which has $T^2$ with modular parameter \(\tau\) as its boundary. It is the BTZ black hole, which
is described by the Euclidean line element
$$
ds^2 = N^2(r) dt_E^2
+\ell^{-2}N^{-2}(r)dr^2+r^2(d\phi+N_\phi(r)dt_E)^2
$$
with
$$
N^2(r)={(r^2-\tau_2^2)(r^2+\tau_1^2)\over r^2},\qquad\quad
N_\phi(r)={\tau_1\tau_2\over r^2}\;.
$$
Here $\tau=\tau_1+i\tau_2$ is the modular parameter of the
boundary torus. Using (\ref{Gc}), one can compute the Euclidean action of this solution
and obtain \cite{Maldacena:1998bw}
$$
S=-{\pi c\over6 } {\rm Im}{1\over \tau}\;.
$$

For the present purpose of counting BPS states, one needs to
consider extremal BTZ black holes. With the Minkowski signature
this means that its mass and angular momentum are equal. After
analytic continuation to a Euclidean complexified geometry, one
finds that the action has become complex and equals $i\pi {c\over
12}\tau$.

Note that a torus with modular parameter $\tau$ is equivalent to a
torus with parameter $a\tau+b\over c\tau+d$, since they differ
only by a relabelling of the $A$- and $B$-cycles. But the
Euclidean BTZ solution labelled by $a\tau+b\over c\tau+d$ in
general differs from the one labelled by $\tau$, with the
difference being that these three-dimensional geometries fill up
the boundary torus in distinct ways. Namely, for the above BTZ
solution the torus is filled in such a way that its $A$-cycle is
contractible. After a modular transformation, this would become
the $\gamma(A)=cA+dB$ cycle. In fact, the BTZ black hole is
related to thermal $AdS_3$ with metric
\be
ds^2 =(r^2+\ell^2)dt_E^2+{dr^2\over r^2+\ell^2} +r^2d\phi^2
\ee
after interchanging the $A$- and $B$-cycles and with $t_E$ and \(\phi\) periodically identified as
$$
t_E\equiv t_E+ 2\pi n \tau_2\qquad, \qquad \phi\equiv \phi+2\pi n\tau_1\;.
$$
In this case the $B$-cycle is non-contractible, while the $A$-cycle is now contractible.
Notice that in this metric it is manifest that $\tau\to\tau+1$ gives
the same geometry. The Euclidean action for this geometry is $S=i\pi {c\over 12}\tau$.


\begin{figure}[htb!]
\centering%
\includegraphics[height=4.5cm]{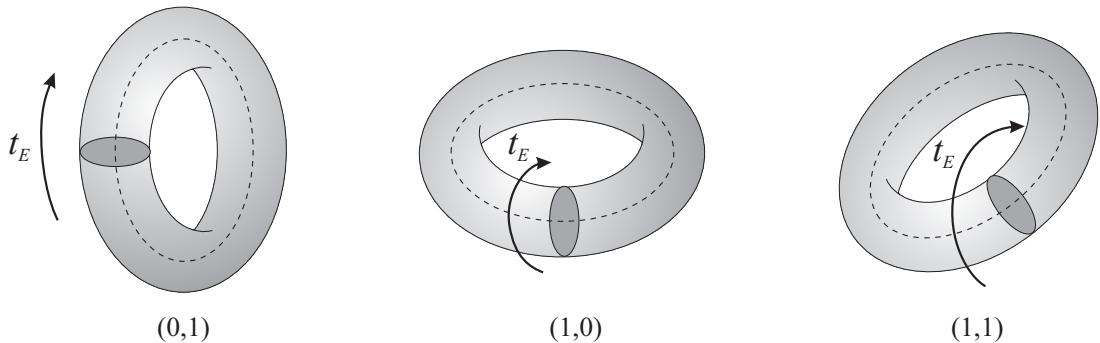}
\caption{\small{(i) Thermal \(AdS_3\), with the B-cycle being
contractible. (ii) BTZ black hole, with the A-cycle being
contractible. (iii) The geometry with the (A+B)-cycle being
contractible.}} \label{fig:Farey}
\end{figure}

The classical geometries with a given boundary torus with modular
parameter $\tau$ can now be obtained from either the thermal AdS
background or the BTZ background by modular transformations. For
definiteness let us take the thermal AdS as our reference point,
so that the classical action for the geometry obtained by acting
with an element $\gamma$ of the modular group is
$$
S=i\pi{c\over 12}\left({a\tau+b\over c\tau+d}\right)\;.
$$
One easily recognizes that these solutions precisely give all  the
leading contributions in the Farey tail expansion corresponding to
$n=0$ terms in (\ref{Zexpansion}). In fact, these terms occur with
a multiplicity one since they represent the vacuum of the SCFT.
The other terms in the expansion should then be regarded as
dressing the Euclidean background with certain contributions that
change the energy of the vacuum. Already in the original Farey
tail paper these contributions were interpreted as coming from
certain (virtual) particles that circle around the
non-contractible cycle. In fact, in the next section we will give
a further justification of this interpretation for the case of the
attractor black holes. Specifically, by using the arguments of
Gaiotto et al.\cite{Gaiotto:2006ns}, we find that the subleading
contributions are due to a gas of wrapped M2-branes which carry
quantum numbers corresponding to the charges and the spin in the
$AdS_3$ geometry. The truncation to the polar terms can in turn be
interpreted as imposing the restriction that the gas of particles
are not heavy enough yet to form a black hole. The $AdS_3$
geometry carries a certain negative energy which allows a certain
amount of particles to be present without causing gravitational
collapse. However, when the energy surpasses a certain bound then
a black hole will form through a Hawking-Page transition. In the
case of the $AdS_3/CFT_2$ correspondence such an interpretation
was first proposed by Martinec \cite{Martinec:1998wm}. Indeed,
when one compares the elliptic genus computed from supergravity to
the exact CFT answer, one finds a mismatch at the threshold for
black hole formation \cite{db}, see also \cite{Dijkgraaf:2000fq}.

The only feature of the Rademacher formula that really depends on
the system at hand is the theta function. The origin of the theta
function lies in the chiral bosons on the world sheet of the
string reduction of the M5-brane. From a spacetime perspective,
however, these theta functions are due to the gauge fields and in
particular have their origin in the Chern-Simons term as discussed
in section~2. In fact, the partition function of a spacetime
effective theory that includes precisely such Chern-Simons terms
in addition to the usual Yang-Mills action was analyzed in detail
by Gukov et al.  in \cite{Gukov:2004id}. These authors showed that
the partition function indeed decomposes in to a sum of
Siegel-Narain theta functions. The $\taubar$-dependence arises
because one of the gauge field components is treated differently
from the others, to ensure that the partition function indeed
converges. Formally it is possible to treat all gauge fields in
the same way, and choose boundary conditions so that the partition
function is holomorphic in $\tau$. This indeed produces the
correct theta functions that we used above. In this paper we will
not give further details of this calculation. A recent discussion
in which parts of this calculation were carefully worked out is
\cite{Kraus:2006nb}.

\subsection{Interpretation in terms of M2-branes}

In the previous subsection we explained the origin of the
exponential factors and the theta function. It is an interesting
question whether one can also say more about the degeneracies
$d_\mu(\hat{q}_0)$ with $\hat{q}_0<0$ that appear in the truncated
characters $\chi_\mu^-(\tau)$. In particular one would like to
give a more detailed accounting of those states from the point of
view of string theory on CY$\times S^2 \times AdS_3$.  In fact, a
nice physical picture of a large class of these states was given
by Gaiotto et al. in terms of M2 and anti-M2 branes which fill up
Landau levels near the north and south pole of the $S^2$
respectively \cite{Gaiotto:2006ns}. In a dilute gas approximation,
this gives rise to the following contribution to the index $Z$:
\be
\label{sugra1} Z_{\rm gas}(\tau,y) = e^{-2\pi i c_L \tau/24}
Z_{\rm sugra}(\tau) Z'_{GV}(\tau, {\textstyle{\frac{1}{2}}} p \tau
+ y) Z'_{GV}(\tau, {\textstyle{\frac{1}{2}}} p \tau - y)
\ee
where $Z_{sugra}(\tau)$ is the contribution from supergravity
modes which is equal to $\prod_{n>0} (1-e^{2\pi i n\tau})^{-n
\chi_{CY}}$,\footnote{A more detailed discussion which corrects
this expression can be found in \cite{Kraus:2006nb}.} and
$Z'_{GV}$ is the (reduced) Gopakumar-Vafa partition function
\be
\label{gv} Z'_{GV}(\tau,y) = \prod_{q,m} \left[ \prod_{n\geq 0}
\left( 1 - e^{2\pi i \tau(m+n) + 2\pi i y q
}\right)^n\right]^{N^m_q}
\ee
where we only take into account contributions with $q\neq 0$. From
the point of view of the world sheet SCFT, the M2/anti-M2 brane
gas describes a collection of states that is freely generated by a
collection of chiral vertex operators. It is clear that
(\ref{sugra1}) suffers from all kinds of limitations. The dilute
gas approximation will eventually break down, there could be other
BPS configurations that contribute, the Landau levels can start to
fill out the entire $S^2$, the $SU(2)$ quantum numbers are bounded
by the level of the $SU(2)$ current algebra, etc. Furthermore, it
also does not exhibit the right behavior under spectral flow.

But suppose the above expression is correct in the low temperature
regime and for small M2-brane charges.
Furthermore let us focus on the regime $\qhat <0$ so that no black
hole shall be formed, so that we are only interested in the states
counted by the truncated characters $\chi^-_\mu(\tau)$. Then we
can make the following identification
\be
\
 \sum_\mu\chi^-_\mu(\tau)e^{i\pi\tau\mu^2+2\pi i \mu\cdot y} \sim \Bigl[Z_{\rm gas}(\tau,y)\Bigr]_{trunc}
\ee
where the r.h.s. is given by (\ref{sugra1}) but truncated in two
ways: it does not only contain terms for which $\hat{q}_0<0$ but
also the total M2-brane charge is restricted to those values for
which the dilute gas approximation is valid, for which the $SU(2)$
quantum numbers do not exceed the unitarity bounds of the CFT,
etc. The most stringent interpretation of the truncation would be
to consider only M2-brane charges that are restricted to lie in
some fundamental domain under spectral flow, that is $q\in
\Lambda^*/\Lambda+ \frac{p}{2}$. However, we will leave the
precise nature of the truncation deliberately vague, and leave a
more precise understanding to future work.

In  \cite{Gaiotto:2006ns} it was shown that the un-truncated form
of this expression together with the modularity of the elliptic
genus gives a derivation of the OSV conjecture. We will now
reexamine these arguments from the point of view of the Farey tail
expansion, while also taking into account the spectral flow
invariance. First notice that the spectral flow invariance can be
restored by extracting $\chi^-_\mu(\tau)$ from this expression,
and using the theta functions to build the full truncated elliptic
genus. After a small calculation one obtains
\bea \label{trunc}
 \sum_\mu\chi^-_\mu(\tau)\theta_\mu(\tau,y)& \sim &\\ \nonumber & & \mbox{${}$}
 \!\!\!\!\!\!\!\!\!\!\!\!\!\!\!\!\!\!\!\!\!\!\!\!\!\!\!\!\!\!\!
 \!\!\!\!\!\!\!\!\!\!\!\!\!\!\!\!\!\!\!\!\!\!\!\!\!\!\!\! e^{-2\pi i c_L
\tau/24}  \sum_{k\in \Lambda} e^{i\pi \tau k^2 + 2\pi i y k}e^{\pi
i p k} \Bigl[Z_{\rm sugra}(\tau) Z'_{GV}(\tau,
{\textstyle{\frac{1}{2}}} p \tau + y + k\tau) Z'_{GV} (\tau,
{\textstyle{\frac{1}{2}}}  p \tau - y - k\tau).\Bigr]_{trunc}
\eea
To obtain the Farey tail expansion of the full partition function
we have to sum over the images under modular group. This is most
easily done after first rewriting the r.h.s. by observing that
\bea -\frac{c_L\tau}{24} + \frac{\tau}{2} k^2 + yk
+\frac{y^2}{2\tau}& =&  -\frac{1}{6\tau^2} \left(
({\textstyle{\frac{1}{2}}}  p \tau + y + k\tau)^3 +
({\textstyle{\frac{1}{2}}}  p \tau - y- k \tau)^3 \right)
\nonumber
\\ & & -\frac{1}{24} c_2 \cdot \left(({\textstyle{\frac{1}{2}}} p \tau + y +
k\tau) + ({\textstyle{\frac{1}{2}}}  p \tau - y- k \tau) \right),
\eea where we used on the l.h.s. $d_{ab}$ to define the quadratic terms, while the cubic terms on the r.h.s. are defined with the help of $d_{abc}$.
Here we recognize precisely the perturbative genus zero and genus
one piece of the topological string partition function. If we
therefore define \be Z_{\rm top}(\tau,y) = Z^{\frac{1}{2}}_{\rm
sugra}(\tau) e^{- 2\pi i (\frac{1}{6\tau^2} y^3 + \frac{c_2
y}{24}) } Z'_{GV}(\tau,y) \ee we can rewrite (\ref{trunc}) as
\be \label{sugra3}
 \sum_\mu\chi^-_\mu(\tau)\theta_\mu(\tau,y) \sim e^{-\frac{\pi i y^2}{\tau}}
 \sum_{k\in \Lambda} e^{\pi i p k}\Bigl[ Z_{\rm
top }(\tau, {\textstyle{\frac{1}{2}}} p\tau + y + k\tau) Z_{\rm
top} (\tau, {\textstyle{\frac{1}{2}}} p\tau - y
-k\tau)\Bigr]_{trunc}.
\ee The full partition function $Z(\tau,y)$ is equal to the sum
over the modular images of (the Farey tail transform of)
(\ref{sugra3}).

To find the entropy one has to extract the leading behavior as
${\rm Im}\,\tau \rightarrow 0$. There is a single term which
dominates, which is the $\tau\rightarrow -1/\tau$ transform of
(\ref{sugra3}). Further, the leading behavior is determined mainly
by the genus zero and genus one contributions. The subleading,
exponentially suppressed "tail" contributions are a representation
of the higher genus contributions. The fact that this sum is
truncated is not noticeable in perturbation theory, since it
hardly affects the entropy calculation. Notice that all bounds on
the M2-brane charges disappear for large M5 charge $p^a$, and
hence in this regime we can ignore the truncation all together.

So we can approximate the full partition function by its leading term in the Farey tail expansion
 \be Z(\tau,y) \sim \sum_{k\in \Lambda} e^{\pi i p k} Z_{\rm top
}\left(-\frac{1}{\tau}, \frac{{{-\frac{1}{2}}}  p  + y -
k}{\tau}\right) Z_{\rm top} \left(-\frac{1}{\tau},-
\frac{{{-\frac{1}{2}}} p  - y + k}{\tau}\right). \ee where we
removed the truncation,  ignored some powers of $\tau$ and used
the fact that the prefactor in (\ref{sugra3}) precisely cancels
under the $\tau\to -1/\tau$ transformation.

To get from here to the OSV conjecture for the entropy it is even
nicer to keep the sum over $d$ in the Farey Tail expansion as
well, which results in \bea \label{sugra4} Z(\tau,y) & \sim  &
\sum_{d\in\mathbb Z}\sum_{k\in \Lambda} e^{\pi i p k} Z_{\rm top
}\left(-\frac{1}{\tau+d}, \frac{{{-\frac{1}{2}}}  p  + y -
k}{\tau+ d}\right) \nonumber
\\ & & \times Z_{\rm top} \left(-\frac{1}{\tau+ d},
\frac{{{-\frac{1}{2}}}  p  - y + k }{\tau+ d}\right)
\eea The entropy $\Omega(p,q)$ can now be written as a multiple contour
integrals of
\be \label{aux11}
Z(\tau,y) e^{-2\pi i (q_0 \tau+ q_a y^a)}.
\ee
Crucially, this is a properly periodic function of $y$. The
peculiar phase $e^{\pi i p k}$ in (\ref{sugra4}) cancels against a
similar phase coming from $e^{-2\pi i (q_0 \tau+ q_a y^a)}$,
because the membrane charge lattice is shifted by $p/2$.
Therefore, the contour integrals together with the sum over $k\in
\Lambda$ and the sum over $d$ can be rewritten as integrals over
the entire imaginary axis, so that at the end of the day the
entropy becomes an inverse Laplace transform of $|Z_{\rm top}|^2$,
which is precisely the OSV conjecture.

Notice that we have been somewhat inaccurate in keeping track of
the $p/2$ shift in part of the above computation, as on the right
hand side of (\ref{trunc}) the membrane charge lattice is centered
around zero rather than $p/2$. The correct expression involves a
sum $\sum_{k\in \Lambda+p/2}$ rather than $\sum_{k\in\Lambda}$.
This has little effect, as one can at the end, in (\ref{sugra4}),
undo this shift by a shift of $y$, which in turn generates an
extra phase $e^{\pi i p q}$ in (\ref{aux11}). This phase should
indeed be present, as the definition (\ref{part}) of the
generalized elliptic genus did precisely involve such an extra
phase.

It is worth emphasizing that the occurence of the topological
string in the degeneracies of the M2 and anti-M2 brane gas (\`a la
Gopakumar-Vafa) and the OSV-conjecture are naturally related by
the Farey tail expansion. In fact, each term in the Farey tail
expansion has a representation in terms of a square of a
topological string partition function with coupling constant equal
to $(a\tau+b)/(c\tau+d)$.  This representation is however
approximate because it assumes a complete decoupling of the M2 and
anti-M2 brane states, which clearly breaks down for large
M2-charges or for small M5-brane charge. We expect that a more
complete analysis will involve corrections in a way similar to the
ones found for 2d Yang-Mills  \cite{Dijkgraaf:2005bp}. We hope to
come back to this point in a future paper.

\section{Discussion}

In this paper we present the Farey tail expansion for \(N=2\)
attractor black holes. The central idea of this expansion is that
the one has to first truncate the partition function so that it
includes only particular low energy states, and then sum over all
images of it under the modular group. Each term can be interpreted
as representing the contribution of a particular (semi-)classical
background. The formula can thus be regarded as partly microscopic
(as the states counted in the "tail") as well as macroscopic (as
the sum over classical backgrounds). We would like to emphasize
that in this expansion, there is no one to one correspondence
between microstates and gravitational backgrounds (as suggested by
Mathur et al, see e.g. \cite{Mathur}). Quite on the contrary, the
major part of the entropy is carried by one particular black hole
background.

The supergravity interpretation of the Farey expansion involves a
natural complete collection of backgrounds of a given type. It is
natural to ask whether the expansion can be refined by including
more general macroscopic backgrounds. It is indeed likely that
such refinements exist, but one expects that these will follow a
similar pattern: one has to truncate the microscopic spectrum even
further and replace the contribution of the omitted states by
certain classical backgrounds. Here one can think of various type
of backgrounds, such as multi-centered solutions ("baby
universes"), bubbling solutions that deform the horizon geometry,
black rings...etc.  A large class of such solutions is known, but
the list is presumably incomplete, and it remains an interesting
problem to use them in a systematic or let alone exact manner.

\section*{Acknowledgements}

We like to thank Frederik Denef, Finn Larsen, Boris Pioline, Andy
Strominger, Cumrun Vafa and Xi Yin for useful discussions. This
research is supported financially by the Foundation of Fundamental
Research on Matter (FOM) and a Spinoza Grant of the Dutch Science
Organization (NWO).

\appendix

\section{Restoring the $\taubar$-dependence}\label{appendix}
\setcounter{equation}{0}

In our discussion of the elliptic genus, from equation
(\ref{notaubar}) onwards we put \(\tau=\bar{\tau}\) since we
argued that the \(\taubar\)-dependence can always be reconstructed
by using (\ref{heateq}) and thus we do not need an extra potential
\(\taubar\) to label the charges of the supersymmetric states.
This argument, though, is only formal in the sense that we need
the \(\taubar\)-dependence for the sum to converge. For
completeness, we present the computation with $\taubar$-dependence
restored in this appendix.

We begin with the counterpart of (\ref{notaubar}) but now with
proper \(\taubar\) dependence and also the $e^{\pi i p\cdot q}$ included
\begin{equation}
Z_{p^a}(\tau,\taubar,y_L, y_R)=\sum_{\{q_0, q\in
  \Lambda^\ast+p/2\}} d(q_0, q)e^{\pi i p\cdot q}e^{2\pi i\tau q_0} e^{i\pi(\tau- \taubar) q_R^2 } e^{2\pi i y\cdot q}.
\end{equation}
 The spectral flow argument for the BPS states implies again
\be
d(n+\frac{1}{2}q_L^2,q_L)=d_\mu(n) \;,
\ee
with $q=\mu\,\, \mathrm{mod}\,\, \Lambda $, $\mu \in
\Lambda^*/\Lambda +p/2$. Therefore we can decompose the elliptic
genus in holomorphic modular forms and non-holomorphic theta
functions:
\be \label{decomA3}
Z_{p^a}(\tau,\bar{\tau},y_L,y_R) =
\sum_{\mu\in\Lambda^{\ast}/\Lambda+\frac{p}{2}}
 \chi_{\mu} (\tau) \theta_{\mu} (\tau,\bar{\tau},y^a),
 \ee
 with
 \bea
 \chi_\mu(\tau) &= &\sum_{-\frac{c_L}{24} \leq n} d_\mu(n) e^{2 \pi i \tau n},\\ \nonumber
 \theta_\mu(\tau,\taubar, y_L, y_R) & = & \sum_{k\in \Lambda}
 e^{\pi i p\cdot \mu} e^{\pi i p\cdot k} e^{2 \pi i \tau\frac{(\mu_L+k_L)^2}{2} } \,e^{-2 \pi i \taubar\frac{(\mu_R+k_R)^2}{2} }\\
&& \times \,e^{2 \pi i y_L\cdot(\mu_L+k_L)} \,e^{2 \pi i y_R\cdot(\mu_R+k_R)}
 \eea


The transformation properties of these theta-functions can be
calculated straight forwardly \cite{borcherds-1998-132}. First of all
they satisfy
\bea \nonumber
\theta_\mu(\tau+1,\taubar+1, y_L,y_R)
& = & \sum_{k\in \Lambda}
 e^{\pi i p\cdot \mu} e^{\pi i p\cdot k} e^{\pi i (\mu+k)^2} \,e^{2 \pi i \tau\frac{(\mu_L+k_L)^2}{2} } \,e^{-2 \pi i \taubar\frac{(\mu_R+k_R)^2}{2} }\\
&& \times \,e^{2 \pi i y_L\cdot(\mu_L+k_L)} \,e^{2 \pi i y_R\cdot(\mu_R+k_R)}\\
&=&e^{\pi i
\mu^2}\,\theta_\mu(\tau,\taubar,  y_L,y_R).
\eea

The presence of the Freed-Witten anomaly, that is, the fact that
\(\mu'\equiv \mu -\frac{p}{2} \in \Lambda^\ast/\Lambda\), is
essential for the above equation to hold: since \(k\in \Lambda\),
we can think of the M2 brane represented by \(k\) as wrapping an
integral two-cycle embedded inside the divisor \({\cal{P}}\). In
this case one can apply the adjunction formula
\be Q\cdot Q+Q \cdot [{\cal{P}}] = 2g-2\;\;\; \Rightarrow\;\;k^2 = k\cdot p \mbox{   mod  } 2\;,
\ee
and exploit the relation
\bea \nonumber
(\mu+k)^2 &=& k^2 + 2(\mu'+\frac{p}{2})\cdot k + (\mu'+\frac{p}{2})^2\\ \nonumber
&=& k^2 + k\cdot p + 2 \mu'\cdot k  + (\mu'+\frac{p}{2})^2\\
&=& \mu^2  \mbox{   mod  } 2
\eea
to arrive at the desired transformation property of the theta-function.

Under S-transformation, one has
\bea\nonumber
\theta_\mu(\frac{-1}{\tau},\frac{-1}{\taubar},
  \frac{y_L}{\tau},\frac{y_R}{\taubar})&=&\sum_{k\in\Lambda}
e^{-\pi i p\cdot \mu} e^{-\pi i p\cdot k} e^{-\frac{2 \pi i
}{\tau}\frac{(\mu_L+k_L)^2}{2} } \,e^{\frac{2 \pi i
}{\taubar}\frac{(\mu_R+k_R)^2}{2} } \\ \nonumber
&& \times \,e^{\frac{2 \pi i y_L}{\tau}\cdot(\mu_L+k_L)} \,e^{\frac{2 \pi i y_R}{\taubar}\cdot(\mu_R+k_R)}\\
\nonumber
 &=&\sqrt{\frac{\left|\Lambda\right|}{\left|\Lambda^*\right|}}e^{-\pi
  i p^2/2}e^{\pi
i \left(\frac{y_L^2}{\tau}-\frac{y_R^2}{\taubar}\right)}
\left(-i\tau \right)^{(b_2-1)/2} \left(i\taubar
\right)^{1/2}\times \\
 &&\sum_{\delta\in \Lambda^*/\Lambda}e^{-2\pi i\mu\cdot
\delta}\theta_\delta(\tau,\taubar,y_L,y_R)\;,
 \eea
where we have performed a Poisson resummation. We can see that the theta-functions have weights \((\frac{b_2-1}{2}, \frac{1}{2})\) and from this we can
deduce that the weights of the characters \(\chi_\mu(\tau)\) are \((\frac{-b_2}{2}-1, 0)\).

Notice that all the \(\taubar\)-dependence in
\(Z_{p^a}(\tau,\bar{\tau},y_L,y_R) \) is contained in the theta
function factor as before; in other words, the incorporation of
the \(\taubar\)-dependence does not change the fact that all the
information about the black hole degeneracies is contained in
holomorphic characters \(\chi_\mu(\tau)\). This is of course again
a consequence of the supersymmetry condition. The desired
transformation properties of the elliptic genus require that the
vector valued modular form $\chi_\mu(\tau)$ transforms in a
conjugate representation as the theta function. Its transformation
properties are
\begin{eqnarray}
\chi_\mu(\tau+1)&=&e^{-\pi i \mu^2}\chi_\mu(\tau), \nonumber \\
\chi_\mu(\frac{-1}{\tau})&=&\sqrt{\frac{\left|\Lambda^*\right|}{\left|\Lambda\right|}}
e^{\pi i p^2/2} e^{\pi i (b_2-2)/4}\left(\frac{1}{\tau} \right)^{(b_2+2)/2}\sum_{\delta\in\Lambda^*/\Lambda}e^{2\pi i\mu\cdot \delta}h_\delta(\tau).
\end{eqnarray}

In order to write the elliptic genus as a Rademacher expansion, we
have to perform a Farey transform to $\chi_\mu(\tau)$. The transformed
elliptic genus \(\tilde{Z}_{p^a}(\tau,\taubar,y_L,y_R)\) has
positive weights \(\tilde{w} = \frac{5}{2}+b_2,
\tilde{\bar{w}}=\frac{1}{2}\). \(\tilde{Z}_{p^a}(\tau,
\taubar,y_L,y_R)\) written as a Rademacher expansion reads
\bea\nonumber
\tilde{Z}_{p^a}(\tau, \taubar, y_L,y_R) &= &
 \sum_{ \mu\in \Lambda^{\ast}/\Lambda+\frac{p}{2} }
 \,\sum_{ \gamma\in\Gamma/\Gamma_{\infty} }
 \tilde{\chi}_\mu^- \left( \frac{a\tau+b}{c\tau+d}\right)
 \theta_\mu\left(\frac{a\tau+b}{c\tau+d},\frac{a\taubar+b}{c\taubar+d}, \frac{y_L}{c\tau+d}, \frac{y_R}{c\taubar+d}\right)\\
&& \left({c\tau+d} \right)^{-({5\over 2}+b_2)}\,\left({c\taubar+d} \right)^{-\frac{1}{2}}e^{-\pi i \frac{c \,y_L^2}{c\tau+d} }e^{\pi i \frac{c \,y_R^2}{c\taubar+d} }\;,
\eea
where
\be
\tilde{\chi}_\mu^-(\tau)= \sum_{{-\frac{c_L}{24}} \leq  n < 0} \tilde d_{\mu}(n)\;e^{2\pi i \tau n}
\ee
now contains only the "polar part" as before. This is our convergent
attractor Farey Tail with the explicit \(\taubar\)-dependence.


\begin{thebibliography}{100}


\bibitem{Dijkgraaf:2000fq}
  R.~Dijkgraaf, J.~M.~Maldacena, G.~W.~Moore and E.~P.~Verlinde,  ``A black hole farey tail,''
  [arXiv:hep-th/0005003].


\bibitem{Strominger:1996sh}
  A.~Strominger and C.~Vafa,
  ``Microscopic Origin of the Bekenstein-Hawking Entropy,''
  Phys.\ Lett.\ B {\bf 379}, 99 (1996)
  [arXiv:hep-th/9601029].



\bibitem{Dijkgraaf:1996xw}
   R.~Dijkgraaf, G.~W.~Moore, E.~P.~Verlinde and H.~L.~Verlinde,
``Elliptic genera of symmetric products and second quantized
strings,''
   Commun.\ Math.\ Phys.\  {\bf 185} (1997) 197
   [arXiv:hep-th/9608096].



\bibitem{Maldacena:1997de}
  J.~M.~Maldacena, A.~Strominger and E.~Witten,
   ``Black hole entropy in M-theory,''
  JHEP {\bf 9712}, 002 (1997)
  [arXiv:hep-th/9711053].


\bibitem{LopesCardoso:1998wt}
  G.~Lopes Cardoso, B.~de Wit and T.~Mohaupt,
``Corrections To Macroscopic Supersymmetric Black-Hole Entropy,''
  Phys.\ Lett.\ B {\bf 451} (1999) 309
  [arXiv:hep-th/9812082].
  G.~Lopes Cardoso, B.~de Wit and T.~Mohaupt,
  ``Deviations from the area law for supersymmetric black holes,''
  Fortsch.\ Phys.\  {\bf 48} (2000) 49
  [arXiv:hep-th/9904005]. G.~Lopes Cardoso, B.~de Wit and T.~Mohaupt,
   ``Macroscopic entropy formulae and non-holomorphic corrections for
  supersymmetric black holes,''
  Nucl.\ Phys.\ B {\bf 567} (2000) 87
  [arXiv:hep-th/9906094].


\bibitem{Ooguri:2004zv}
  H.~Ooguri, A.~Strominger and C.~Vafa,
   ``Black hole attractors and the topological string,''
  %
  Phys.\ Rev.\ D {\bf 70}, 106007 (2004)
  [arXiv:hep-th/0405146].


\bibitem{Gopakumar:1998ii}
  R.~Gopakumar and C.~Vafa,
  ``M-theory and topological strings. I,''
  [arXiv:hep-th/9809187].
  ``M-theory and topological strings. II,''
  [arXiv:hep-th/9812127].

\bibitem{Gaiotto:2006ns}
 D.~Gaiotto, A.~Strominger and X.~Yin,
  ``From AdS(3)/CFT(2) to black holes / topological strings,''
  [arXiv:hep-th/0602046].

\bibitem{Denef-Moore}
F. Denef and G.W. Moore, to appear.


  \bibitem{Gaiotto:2006}
 D.~Gaiotto, A.~Strominger and X.~Yin,
  ``The M5-Brane Elliptic Genus:
Modularity and BPS States,''
  [arXiv:hep-th/0607010].

\bibitem{Kraus:2006nb}
  P.~Kraus and F.~Larsen,
  ``Partition functions and elliptic genera from supergravity,''
  arXiv:hep-th/0607138.

\bibitem{Manschot:2007ha}
  J.~Manschot and G.~W.~Moore,
  ``A Modern Farey Tail,''
  arXiv:0712.0573 [hep-th].





\bibitem{Freed:1999vc}
  D.~S.~Freed and E.~Witten,
  ``Anomalies in string theory with D-branes,''
  [arXiv:hep-th/9907189].

\bibitem{Witten:1996hc}
  E.~Witten,
  ``Five-brane effective action in M-theory,''
  J.\ Geom.\ Phys.\  {\bf 22}, 103 (1997)
  [arXiv:hep-th/9610234].




\bibitem{Minasian:1999qn}
  R.~Minasian, G.~W.~Moore and D.~Tsimpis,
  ``Calabi-Yau black holes and (0,4) sigma models,''
  Commun.\ Math.\ Phys.\  {\bf 209}, 325 (2000)
  [arXiv:hep-th/9904217].

\bibitem{zagier}
M. Eichler and D. Zagier, {\it The theory of Jacobi forms},
Birkh\"auser 1985.




\bibitem{Maldacena:1999bp}
  J.~M.~Maldacena, G.~W.~Moore and A.~Strominger,
   ``Counting BPS black holes in toroidal type II string theory,''
  [arXiv:hep-th/9903163].




\bibitem{Maldacena:1998bw}
  J.~M.~Maldacena and A.~Strominger,
  ``AdS(3) black holes and a stringy exclusion principle,''
  JHEP {\bf 9812}, 005 (1998)
  [arXiv:hep-th/9804085].

\bibitem{Martinec:1998wm}
  E.~J.~Martinec,
  ``Conformal field theory, geometry, and entropy,''
 [arXiv:hep-th/9809021].

\bibitem{db}
J.~de Boer,
  ``Large N Elliptic Genus and AdS/CFT Correspondence,''
  JHEP {\bf 9905}, 017 (1999)
  [arXiv:hep-th/9812240].

\bibitem{Gukov:2004id}
  S.~Gukov, E.~Martinec, G.~W.~Moore and A.~Strominger,
  ``Chern-Simons gauge theory and the AdS(3)/CFT(2) correspondence,''
  arXiv:hep-th/0403225.


\bibitem{Dijkgraaf:2005bp}
  R.~Dijkgraaf, R.~Gopakumar, H.~Ooguri and C.~Vafa,
  ``Baby universes in string theory,''
  Phys.\ Rev.\ D {\bf 73} (2006) 066002
  [arXiv:hep-th/0504221].


\bibitem{Mathur}
  S.~D.~Mathur,
 ``The fuzzball proposal for black holes: An elementary review,''
  Fortsch.\ Phys.\  {\bf 53} (2005) 793
  [arXiv:hep-th/0502050], and the references therein.


\bibitem{borcherds-1998-132}
 R.~E.~Borcherds,
``Automorphic forms with singularities on Grassmannians,''
 Inventiones\ mathematicae {\bf 132}, 491 {1998}
 [arXiv:alg-geom/9609022]





\end{thebibliography}

\end{document}